\documentclass[11pt,preprint]{aastex}

\newcommand{\HII}{H {\small II}  }
\newcommand{\kms}{{\rm ~km~s}^{-1}}

\newcommand{\tothe}{^}

\slugcomment{To appear in ApJ}

\shorttitle{RARE METHANOL MASERS IN NGC 7538 IRS 1}
\shortauthors{GALV\'AN-MADRID ET AL.}

\begin{document}

\title{THE RARE 23.1-GHZ METHANOL \\ MASERS IN NGC 7538 IRS 1}


\author{Roberto Galv\'an-Madrid\altaffilmark{1,2,3}, 
Gabriela Montes\altaffilmark{2,4}, Edgar A. Ram{\'\i}rez\altaffilmark{2,5}, \\
Stan Kurtz\altaffilmark{2}, Esteban Araya\altaffilmark{6}, and 
Peter Hofner\altaffilmark{7,8,9}
}

\email{rgalvan@cfa.harvard.edu}

\altaffiltext{1}{Harvard-Smithsonian Center for Astrophysics, 60 Garden Street, 
Cambridge MA 02138, USA}
\altaffiltext{2}{Centro de Radioastronom{\'\i}a y Astrof{\'\i}sica, 
Universidad Nacional Aut\'onoma de M\'exico, Morelia 58090, M\'exico}
\altaffiltext{3}{Academia Sinica Institute of Astronomy and Astrophysics, 
P.O. Box 23-141, Taipei 106, Taiwan}
\altaffiltext{4}{Instituto de Astrof{\'\i}sica de Andaluc{\'\i}a, CSIC, 
Camino Bajo de Hu\'etor 50, E-18008 Granada, Espa\~{n}a}
\altaffiltext{5}{Department of Physics and Astronomy, University of Sheffield, 
Sheffield S3 7RH, England}
\altaffiltext{6}{Physics Department, Western Illinois University,
1 University Circle, Macomb IL 61455, USA}
\altaffiltext{7}{Physics Department, New Mexico Tech, 801 Leroy Place, 
Socorro NM 87801, USA}
\altaffiltext{8}{National Radio Astronomy Observatory, P.O. Box 0, 
Socorro NM 87801, USA}
\altaffiltext{9}{Max-Planck-Institut f\"{u}r Radioastronomie, Auf dem H\"{u}gel 69,53121, Bonn, 
Germany}

\begin{abstract}

We present high angular resolution ($\theta_{syn} \lesssim
0\rlap.{''}2$) observations of the 23.1-GHz methanol (CH$_3$OH)
transition toward the massive star forming region NGC 7538 IRS 1. The
two velocity components previously reported by Wilson et al. are
resolved into distinct spatial features with brightness temperatures
($T_B$) greater than $10^4$ K, proving their maser nature. Thus, NGC
7538 IRS 1 is the third region confirmed to show methanol maser
emission at this frequency.  The brighter 23.1-GHz spot coincides in
position with a rare formaldehyde (H$_2$CO) maser, and marginally with a 22.2-GHz 
water (H$_2$O) maser, for which we report archival observations. 
The weaker CH$_3$OH spot coincides with an H$_2$O maser. 
The ratio of $T_B$ for the 23.1-GHz masers to
that of the well-known 12.2-GHz CH$_3$OH masers in this region roughly
agrees with model predictions. However, the 23.1-GHz spots are offset
in position from the CH$_3$OH masers at other frequencies. This is
difficult to interpret in terms of models that assume that all the
masers arise from the same clumps, but it may result from turbulent
conditions within the gas or rapid variations in the background radiation 
field.  \end{abstract}


\section{Introduction}
\label{sec:intro}

NGC 7538 is a Galactic star forming region located 
at a distance of $2.65^{+0.12}_{-0.11}$ kpc
\citep{Mosca09}. Multiple infrared sources were discovered in the
vicinity by \cite{Wynn74}. The brightest of these, IRS~1, was recently 
resolved in the near-IR and mid-IR by \cite{Kraus06} and
\cite{DBM05}, respectively. The centimeter free-free emission 
has a bipolar structure \citep{Camp84,Sandell09,Zhu10} and shows variability \citep{Ramiro04}.
Also, the recombination lines from the ionized gas are 
unusually broad \citep{Gaume95,Sewi04,KZK08}. 

NGC 7538 IRS~1 is an exceptionally rich maser source. Maser emission
has been detected in hydroxyl (OH), water (H$_2$O), ammonia (NH$_3$),
methanol (CH$_3$OH), and formaldehyde (H$_2$CO) (e.g., Gaume et
al. 1991, hereafter G91; Hutawarakorn \& Cohen 2003; Hoffman et
al. 2003, hereafter H03; Kurtz et al. 2004). CH$_3$OH emission at
23.1~GHz, from the $9_2-10_1$ $A^+$ transition, was reported for NGC
7538 by \cite{Wil84} (hereafter W84). They speculated that the
emission was maser in nature, based on observations taken with the
Effelsberg radio telescope ($\theta_{beam} \approx 43\arcsec$). The
well-known 6.7-GHz ($5_1-6_0$ $A\tothe +$) and 12.2-GHz ($2_0-3_{-1}$
$E$) class II CH$_3$OH masers have also been detected toward this
source (Minier et al. 2000, hereafter M00; Minier et al. 2002; Moscadelli et al. 2009, 
hereafter M09). Evidence for a
circumstellar disk in IRS~1, based on a velocity gradient within a
linear structure of 6.7-GHz and 12.2-GHz methanol maser spots, was
presented by \cite{Pesta04, Pesta09}
\citep[see however][]{DBM05}. The warm molecular gas surrounding the hypercompact \HII
region (IRS 1) also has dynamics indicative of rotation
\citep{Klaass09}.

The 23.1-GHz class II CH$_3$OH maser is quite rare.  Until now, only
two regions harboring these masers have been confirmed: W3(OH) and NGC
6334 F (W84; Menten et al. 1985; Menten et al. 1988; Menten \& Batrla 1989).
\cite{Cragg04} observed 50 southern star-forming regions and detected
23.1-GHz maser emission in only one --- the previously known NGC 6334
F.  The 4.8-GHz ($1_{10}-1_{11}$) H$_2$CO maser in NGC 7538 IRS 1 is
also quite rare; at present, only seven of these masers have been
found in the Galaxy \citep[][and references therein]{Araya08}.  And
NH$_3$, although ubiquitous in massive star forming regions, is not a
common maser. Various transitions and isotopes are known to present
maser emission \citep[e.g.,][]{Schilke91,Hofn94,GM09}, but to
date, the metastable $^{15}$NH$_3$ (3,3) maser has only been found in
NGC 7538 IRS~1 \citep[G91;][]{Mauer86,John89,WW96}.

Because the original 23.1-GHz CH$_3$OH detection by W84 was never
pursued at higher angular resolution, and because of the presence
in IRS~1 of not one but possibly three rare maser species, we observed NGC
7538 IRS~1 with the goals of confirming the maser nature of the 23.1-GHz 
CH$_3$OH emission, and accurately locating this emission with
respect to the other masers.  In \S ~2 we describe the observations
and the data reduction procedure. We present the results in \S ~3 and
provide a discussion in \S ~4.

\section{Observations}

\label{sec:obs}

We observed the CH$_3$OH $9_2-10_1$ $A^+$ transition at 1.3 cm
\citep[$\nu_0=23.121024$~GHz,][]{Meh85} toward NGC 7538 IRS~1 with the
VLA of NRAO.\footnote{The National Radio Astronomy Observatory is a
facility of the National Science Foundation operated under cooperative
agreement by Associated Universities, Inc.}  The observations were
made on 2006 May 29 and May 30 for a total on-source time of $\sim 1$
hour (program AG722). The array was in the BnA configuration and the pointing center
was $\alpha\mathrm{(J2000)} = 23^{\mathrm h}$~$13^{\mathrm
m}$~$46\rlap.{^{\mathrm s}}00$, $\delta\mathrm{(J2000)} =
61^{\circ}$~$28^\prime$~$11\rlap.{''}0$.  We used the 1A correlator
mode, measuring right circular polarization without Hanning smoothing.
A 3.125 MHz bandwidth with 255 channels of 12.207 kHz each, provided a
velocity coverage of 40 km s$^{-1}$ and a channel width of 0.16 km
s$^{-1}$.

The data were reduced following standard spectral line procedures with
the AIPS software of NRAO. 
The flux, phase, and bandpass calibrators were J0137+331 ($S_\nu = 1.08$
Jy, calculated using the 1999.2 VLA values), J2322+509 ($S_\nu = 0.93 
\pm 0.02$ Jy, bootstrapped flux density), and J0319+415 ($S_\nu = 11.4
\pm 0.2$ Jy, bootstrapped flux density), respectively.
The data received one iteration of phase-only self-calibration. 
A continuum level of about 300 mJy was subtracted
in the $(u,v)$ plane and bandpass calibration was applied.  The
final image cube was made with the ROBUST parameter of AIPS task IMAGR
set to 0, resulting in a synthesized beam of FWHM = $0\rlap.{''}23
\times 0\rlap.{''}12$ at position angle $+70^\circ$.  The per-channel
noise level of the CLEANed image cube was 7~mJy~beam$^{-1}$.

The highest angular resolution observations of H$_2$O masers in
NGC~7538 reported in the literature are the 5--6$''$ resolution data
of Kameya et al. (1990).  However, archival A-configuration VLA data, 
taken on 1999 July 30 under program AS667 (see Sarma et
al. 2002) are available. 
The data were taken in B1950 coordinates, and were
precessed to J2000 before processing.  The precessed pointing center
was $\alpha$ = 23$^{\rm h}$ 13$^{\rm m} 45\rlap.^{\rm s}333, \delta =
61^\circ 28' 10\rlap.{''}59.$ The observations were made in the 4IF
Hanning-smoothed mode,  observing the water maser line at 22.235
GHz, centered at V$_{LSR} = -58$ $\kms$.  A 0.781 MHz bandwidth
with 63 channels was employed, providing a velocity coverage of
10.5~km~s$^{-1}$ with 0.17~km~s$^{-1}$ channels.

The data were reduced using standard high-frequency calibration
procedures. The flux and phase calibrators were B0134+329 ($S_\nu =
1.13$~Jy, calculated from the 1992.2 VLA parameters) and B2320+506
($S_\nu = 1.33 \pm 0.07$ Jy, bootstrapped flux).  No bandpass
calibration was applied.  The data were self-calibrated in phase and 
three fields were imaged with a synthesized beam of 99.6 $\times$ 77.1 
milli-arcsec at a position angle of $+15^\circ$.  Maser-free channels
had a typical rms noise of 20 mJy~beam$^{-1}$.   One field was $1\rlap.{'}3$
from the pointing center; a primary beam correction was applied to this image.

Positional uncertainty arises from the intrinsic uncertainty in the calibrator
position ($<0\rlap.{''}002$ for B2320+506), phase stability and distance from
the phase tracking center, precession of coordinates, 
and the statistical uncertainty of the Gaussian fit.
The latter quantity depends on the signal-to-noise ratio and varies from one
maser to another.  Taking the sources of uncertainty into
account, we report a worst-case absolute positional error for the water masers of 
$0\rlap.{''}05$.

\section{Results}

\subsection{Methanol Masers: Source Sizes and Positions}
\label{sec:sizes}

We detect emission at two sky positions, and denote them by $N$ (for
North) and $S$ (for South). The two features are nominally resolved at
our angular resolution. The emission in the peak channel of each
component was well-fit by a 2-D Gaussian.  The $S$ component has a
deconvolved size of $0\rlap.{''}11 \times 0\rlap.{''}07$ ($\pm
0\rlap.{''}05$) at position angle
$+73\tothe\circ\pm40\tothe\circ$. The $N$ component has a major axis
of $0\rlap.{''}09$ (at position angle
$+49\tothe\circ\pm10\tothe\circ$); it is unresolved in the orthogonal
direction.

The spatial positions of the two spectral components are separated by
about a beamwidth (see Figure 1).  The 2-D Gaussian fits yield J2000
positions of $\alpha_N = 23^{\mathrm h}$~$13^{\mathrm
m}$~$45\rlap.{^{\mathrm s}}358$, $\delta_N =
61^{\circ}$~$28^\prime$~$10\rlap.{''}45$; and $\alpha_S = 23^{\mathrm
h}$~$13^{\mathrm m}$~ $45\rlap.{^{\mathrm s}}365$, $\delta_S =
61^{\circ}$~$28^\prime$~$10\rlap.{''}28$ for the $N$ and $S$ components,
respectively.  
The absolute position of the phase calibrator (J2322+509) is reported
with an accuracy of $0.002''$ in the VLA database. 
The phase noise in this quasar
after calibration was 
$\lesssim 10^\circ$, and no change in the position of the maser spots was 
introduced by the self-calibration. 
We estimate the absolute positional uncertainty to be not greater than $0\rlap.{''}05$.

\subsection{Methanol Masers: Spectral Line Parameters}
\label{sec:pos}

We fit the spectrum of each emission spot with a single Gaussian
profile (see Fig. 2), to obtain the line-center velocities,
linewidths, and peak intensities listed in Table 1. 
These results are roughly consistent with those of W84. 
The flux densities of the
spectral features are lower by 26 \% and 54 \% (for components N and
S) than in the Effelsberg observations (W84). 
Also, the linewidths at half power in our data are 32 \% smaller (N) and 
52 \% larger (S) than in W84. The centroid velocities match within 0.3 
$\kms$ with those reported by W84. The discrepancies 
could arise for a number of reasons, such as time variability in the masers,  
different Gaussian fitting procedures, and our higher spatial/spectral resolution and 
signal-to-noise ratios.

\subsection{Water Masers}
 
Twenty-one water masers were detected in three distinct clusters.  We
refer to these clusters as M (main; coincident with IRS 1), E (east,
about 6$''$ east of IRS 1 ), and S (south, about 80$''$ south of IRS
1 and corresponding to NGC 7538 IRS 11). The position, peak flux, 
velocity, and velocity range of each maser are reported in Table 2.
Within each cluster the masers are numbered by increasing right
ascension.  The positions of some of the masers corresponding to IRS 1 are
shown in Figure 3.

Water masers frequently occur over a wide velocity range and indeed, maser
emission was found at both extremes of the 0.781 MHz passband. 
Kameya et al. (1990) reported a velocity range of $-45.4$ to
$-83.4$ for water masers around IRS 1--3; it is highly probable
that additional masers are present at velocities outside of
the $-52.8$ to $-62.2$~km~s$^{-1}$ range observed by program AS667.

\section{Discussion}
\label{sec:disc}

\subsection{Nature of the emission}
\label{sec:nature}

W84 did not have sufficient angular resolution to place a useful limit
on the source brightness temperature.  Nevertheless, they argued that
the emission was of maser origin, based on absorption in the
CH$_3$OH~$10_1-9_2 \, A^-$ line.  This absorption implied that
the excitation temperature of the CH$_3$OH~$9_2-10_1 \, A^+$ emission line
was greater than the background temperature of at least 1000~K.
With our $250\times$ higher angular resolution we calculate brightness
temperatures of $T_N>1.3\times10\tothe5$ K and $T_S \sim
4\times10\tothe4$ K for components $N$ and $S$, respectively.  Such high
brightness temperatures cannot reflect the kinetic temperature of the
molecular gas, thus we prove the maser nature of the emission. The
relatively broad linewidths may indicate that the masers are
saturated; alternatively, it could indicate the presence of unresolved maser
components. 

Three objects are now confirmed to harbor 23.1-GHz CH$_3$OH maser emission: 
NGC 7538 IRS 1 (this paper), W3(OH) \citep{Menten85,Menten88}, and NGC 6334F 
\citep{MB89}. 

\subsection{Comparison with other masers}
\label{sec:corr}

Figure 3 shows an overlay of the free-free continuum with the
locations of the masers we discuss here. 
The 6.7-GHz and 12.2-GHz CH$_3$OH masers reported by M00 ({\it right} panel) and the 12.2-GHz 
CH$_3$OH masers reported by M09 ({\it left} panel) are plotted in separate panels for clarity. 
The positions of M00 were shifted by 18 mas to the west and south to account for the 
proper motions reported by M09 for CH$_3$OH masers. 
Although not shown in Figure 3, OH masers reported by Hutawarakorn \&
Cohen (2003) are located slightly south of the IRS~1 core, coincident 
with the more diffuse continuum emission seen at the bottom of Figure 3.  
The most striking feature is the general clustering of the many
different maser species around IRS~1. The bright core of continuum
emission from IRS~1 has linear dimensions of order 1000 AU. Scattered
across this area are the various masers of class II CH$_3$OH,
H$_2$O, NH$_3$, and H$_2$CO. 
There are three spatial coincidences at the present angular
resolutions.  One is the 23.1-GHz component $S$, which coincides with
H$_2$O maser M8.  Another is the 23.1-GHz component $N$, which
coincides with component I of the 4.8-GHz formaldehyde maser (which is itself 
resolved into two components: Ia and Ib, separated by $\sim 60$ AU; H03) 
and is also marginally associated with the H$_2$O maser M7.
The third association is H$_2$O maser M6 with component II
of the H$_2$CO masers.
Table 3 lists the positions, offsets, and absolute positional 
uncertainty of the coincident masers.
The 6.7 or 12.2-GHz CH$_3$OH spot closest to a 23.1-GHz maser is 
component 1 of \cite{Mosca09}, with a separation of 65 mas (or 1.3$\sigma$) 
from our component $N$. As we are confident with our estimates of the
astrometric precision, we still rule out an association between these
features.

The velocities of the 23.1-GHz CH$_3$OH masers ($V_N\approx-56.1$
$\kms$, $V_S\approx-59.1$ $\kms$) are similar to those of the
other masers in the vicinity of IRS~1.  The 6.7-GHz CH$_3$OH masers
have LSR velocities in the range $V_{6.7}=[-61.6,-55.0]$ $\kms$, and
the cross-power spectrum of those masers for the entire field shows 
two velocity components centered at $\approx -56$ $\kms$ and $\approx -58$ $\kms$
(M00). M00 also report several 12.2-GHz CH$_3$OH masers
coincident with 6.7-GHz masers; the velocities of the former are
$\approx -58$ $\kms$. 
M09 reports more CH$_3$OH masers at 12.2 GHz than does M00; all 
of the additional masers are in the velocity interval $[-61.9,-55.8]$ $\kms$.  
The 4.8-GHz H$_2$CO masers also have two
velocity components.  The components Ia and Ib have almost the
same LSR velocity ($V_{Ia} \approx -57.9$ $\kms$, $V_{Ib} \approx
-57.8$ $\kms$), while component II is at $V_{II}\approx -60.2$ $\kms$
(H03). The $\tothe{15}$NH$_3$ (3,3) masers have velocities in the
range $V_{\tothe{15}NH_3}=[-52.8,-61.7]$ $\kms$ (G91).  The H$_2$O 
masers reported here have a velocity range of
$V_{H_2O}=[-53,-63]$ $\kms$ (see Table 1). The typical errors in the 
determination of the velocity features are $\sim 0.1$ $\kms$ or better.

For the positional associations the corresponding velocities are very similar. 
CH$_3$OH maser $S$ is centered at $-59.11$ $\kms$, and H$_2$O maser M8 spans the 
range $[-60.8,-59.2]$ $\kms$. 
CH$_3$OH maser $N$ is centered at $-56.09$ $\kms$ while H$_2$CO 
masers Ia/Ib are at $-57.9/-57.8$ $\kms$. The corresponding H$_2$O maser, M7, 
spans the range $[-58.2,-55.2]$ $\kms$, which includes both the CH$_3$OH and H$_2$CO 
maser velocities. Hence, the maser counterparts coincide both spatially and kinematically.

\subsection{Comparison with models}
\label{sec:models}

Although the 23.1-GHz methanol masers do coincide with water and
formaldehyde masers, they {\it do not} coincide with 6.7-GHz or
12.2-GHz methanol masers. 
Fig. 3 shows that this holds when the positions of the 23.1-GHz CH$_3$OH masers 
are compared to either the masers reported by M00 ($right$) or by M09 ($left$). 
The absolute position uncertainties are $\sim 0\rlap.{"}01$ or better for M00, and 
$\sim 0\rlap.{"}001$ for M09.
It is not surprising that regions 
masing at 6.7 or 12.2 GHz might not show 23.1-GHz maser emission ---
the models of \citet{Cragg05} indicate that the former transitions
mase over a wider range of physical parameters than the latter
transition.  But it {\it is} suprising that 23.1-GHz masers
would not be accompanied by 6.7 or 12.2-GHz masers: if conditions 
are adequate for 23.1-GHz masing, they should also be adequate for 
6.7 and 12.2-GHz masing.  Moreover, current models of methanol masers
\citep{Sobo97a,Sobo97b,Cragg04,Cragg05} predict brightness temperatures 
that are orders of magnitude higher for the 6.7 and 12.2-GHz masers
compared to the 23.1-GHz maser.

In light of current models, then, the puzzle is why the locations
of the 23.1-GHz masers do not also show 6.7 or 12.2-GHz emission. 
A possible explanation is the non-simultaneity of the observations. 
However, the observations of M09 were made within days to months 
of ours, and the maser spots were consistent for all the epochs of 
M09 (Mark Reid, personal communication).

Changing physical conditions as a function of time are to be expected
in the dynamic and turbulent medium of star formation regions \citep[]
[and references therein]{BP07}.  In fact, the turbulent medium may
be intimately related to the production of the masers \citep{Sobo98}.  
Most of these changes will occur on a dynamic timescale, however,
which is typically orders of magnitude longer than astronomical monitoring.
Much more rapid change can be produced in the radiation fields. 
Recent models of \HII regions with accretion activity predict 
the existence of significant variations in the cm free-free flux on  
timescales as short as a few years \citep{Peters10}. Fast variations in the radio flux 
have been reported for IRS1 by \cite{Ramiro04}, which is one of the best 
candidates for an \HII region still undergoing accretion 
\citep{Klaass09,Sandell09,Zhu10}. Moreover, all of the three known 
23.1-GHz CH$_3$OH maser regions have background free-free emission detected at radio wavelengths 
(Menten et al. 1988; Carral et al. 2002; this paper). 
The extended frequency coverage of the Expanded VLA (EVLA) 
will facilitate simultaneous multi-transition maser studies of this type of objects 
at high angular resolution.

Finally, we underscore that if the $N$ 23.1-GHz
maser and the 4.8-GHz H$_2$CO are close in space, then 
similar general conditions may give rise to both rare masers.  
\cite{BdJ81} developed a pumping model for the IRS~1 H$_2$CO maser based on free-free
emission from the \HII region, and \cite{Pratap92} concur
that this model is adequate to explain the 4.8-GHz maser in IRS~1.
Nevertheless, the Boland \& de Jong model also predicts a bright
H$_2$CO maser at 14.5~GHz, and this maser transition was not detected
by \cite{Hoff03}. Furthermore, \cite{Araya07} showed that this model
is not able to explain the H$_2$CO maser in another region of high-mass 
star formation: IRAS 18566+0408. 
We note that the consideration of non-simultaneous observations applies 
equally to these coincident masers as it did to the non-coincident methanol
masers mentioned above.
This caveat notwithstanding, the possible correlation of these two
rare maser species may prove helpful in better defining the pumping
mechanism of each.

\acknowledgements

We thank the anonymous referee for comments that significantly improved this paper. 
We are grateful to the VLA staff for supporting the initial phase of
this project via the NRAO program of observing time for university
classes. We thank R. Franco-Hern\'andez for providing the 2-cm 
image; A. Sobolev, J. Ballesteros-Paredes, and M. Reid for helpful discussions; 
and V. Minier for providing us with maser positions. 
R. G.-M. acknowledges support from an SMA predoctoral fellowship. 
E. A. was partially supported by a NRAO Jansky Fellowship. 
P. H. acknowledges partial support from NSF grant AST-0908901.

\newpage

\begin{deluxetable}{ccc}
\tabletypesize{\scriptsize}
\tablecaption{Spectra Parameters}\label{tab:param}
\tablewidth{0pt}
\tablehead{
\colhead{Parameter} & \colhead{N} & \colhead{S}}
\startdata
 Peak \tablenotemark{a} (mJy beam$^{-1}$) & $331\pm4$ & $97\pm3$ \\
 $T_\mathrm{B,peak}$\tablenotemark{a} (K) & $>1.3\times10\tothe5$ & $\sim 4\times10\tothe4$  \\
 V$_\mathrm{LSR}$\tablenotemark{a} (km s$^{-1}$) & $-56.09\pm0.02$ & $-59.11\pm0.08$ \\
$\Delta V_\mathrm{FWHM}$\tablenotemark{a} (km s$^{-1}$) & $1.78\pm0.04$ & $2.89\pm0.18$ \\
$\theta_\mathrm{S}\tablenotemark{b}$ (arcsec) & $<0.09\times0.09$ & $0.11\times0.07$ 
\enddata
\tablenotetext{a}{From a Gaussian fit to the spectrum of the component. 1$\sigma$ 
statistical errors are quoted.}
\tablenotetext{b}{Spot diameters at half power are from a 2-D Gaussian fit at the peak channel 
of the component, using the task JMFIT in AIPS. For component N 
the value of the minor axis is an upper limit; see \S3.1.}
\end{deluxetable}

\begin{deluxetable}{cccccc}
\tablecolumns{6}
\tabletypesize{\scriptsize}
\tablecaption{Water maser positions and velocities}\label{tab:water}
\tablewidth{0pt}
\tablehead{
\colhead{Maser} & \colhead{R.A. (J2000)} & \colhead{Dec. (J2000)} & \colhead{Peak 
$S_\nu$} &
\colhead{Peak V$_{LSR}$}  & \colhead{Velocity Range} \\
\colhead{Designation\tablenotemark{a}} & \colhead{(sec)\tablenotemark{c}} &  
\colhead{($ '\;''$)\tablenotemark{b}} & \colhead{(Jy)} & \colhead{(km~s$^{-1}$)} &
\colhead{(km~s$^{-1}$)}
}
\startdata
M1  &  45.305  & 28 09.38   &  26.8 &  -61.6  &  -60.8 to -62.3 \\  
M2  &  45.306  & 28 09.33   &  0.90 &  -58.8  &  -58.5 to -59.5 \\  
M3  &  45.324  & 28 10.57   &  31.7 &  -57.8  &  -57.3 to -58.5 \\  
M4  &  45.340  & 28 09.34   &  0.97 &  -62.3  &  -61.5 to -63.1 \\  
M5  &  45.350  & 28 10.42   &  138. &  -60.1  &  -59.5 to -61.1 \\  
M6  &  45.351  & 28 10.39   &  0.43 &  -58.5  &  -58.0 to -59.0 \\  
M7  &  45.362  & 28 10.42   &  0.69 &  -56.7  &  -55.2 to -58.2 \\  
M8  &  45.365  & 28 10.32   &  41.7 &  -60.0  &  -59.2 to -60.8 \\  
M9  &  45.377  & 28 09.59   &  2.15 &  -59.3  &  -58.8 to -60.5 \\  
M10 &  45.383  & 28 10.69   &  0.60 &  -53.4  &  -53.1 to -54.1 \\  
E1  &  46.365  & 28 13.65   &  1.03 &  -55.0  &  -53.4 to -56.2 \\  
E2  &  46.930  & 28 10.17   &  6.91 &  -53.4  &  -53.1 to -54.2 \\  
E3  &  46.956  & 28 09.99   &  0.33 &  -61.9  &  -61.0 to -62.9 \\  
S1  &  44.766  & 26 51.22   &  357. &  -54.9  &  -53.1 to -56.2 \\  
S2  &  44.949  & 26 49.61   &  3.82 &  -57.3  &  -56.8 to -57.8 \\  
S3  &  44.959  & 26 49.73   &  4.47 &  -60.8  &  -60.1 to -62.6 \\  
S4  &  44.964  & 26 49.50   &  20.7 &  -54.1  &  -53.7 to -54.4 \\  
S5  &  44.967  & 26 49.40   &  11.4 &  -56.7  &  -56.4 to -57.2 \\  
S6  &  44.973  & 26 49.43   &  1.49 &  -55.7  &  -54.9 to -56.8 \\  
S7  &  44.973  & 26 49.27   &  2.99 &  -56.2  &  -55.9 to -57.3 \\  
S8  &  44.979  & 26 49.75   &  2.28 &  -62.4  &  -62.1 to -62.8 \\  
\enddata
\tablenotetext{a}{The masers associated with IRS~1 are designated M, those lying
substantially to the east (about $6''$) are designated E, and the cluster to the 
south (corresponding to IRS 11) is designated S. Within each designation the masers
are numbered in order of increasing Right Ascension.}
\tablenotetext{b}{All masers are at 23$^{\rm h}$ 13$^{\rm m}$ R.A.; only seconds are
listed here.  Likewise, all masers are at $+61^\circ$ Dec.; only minutes and seconds
are listed.}
\end{deluxetable}

\newpage

\begin{deluxetable}{cccccc}
\tablecolumns{6}
\tabletypesize{\scriptsize}
\tablecaption{Maser Counterparts}\label{tab:counter}
\tablewidth{0pt}
\tablehead{
\colhead{Maser} & \colhead{R.A. (J2000)} & \colhead{Dec. (J2000)}  & 
\colhead{Abs. Uncertainty} &
\colhead{Offset\tablenotemark{a}} & \colhead{Reference} \\
\colhead{Designation} & \colhead{(h m s)} &  \colhead{($^{\circ}\; '\;''$)} &  
\colhead{($''$)} & \colhead{($''$)} & 
}
\startdata
CH$_3$OH 23.1-GHz $N$ & 23 13 $45.358$ & 61 28 $10.45$  & 0.05 &  --- & This paper \\
H$_2$CO 4.8-GHz $I$\tablenotemark{b} & 23 13 $45.3559$ & 61 28 $10.444$ & 0.003 & 0.02 & Hoffman et al. (2003) \\
H$_2$O $M7$ & 23 13 $45.362$ & 61 28 $10.42$ & 0.05 & 0.04 & This paper \\
\hline
CH$_3$OH 23.1-GHz $S$ & 23 13 $45.365$ & 61 28 $10.28$  & 0.05 & --- & This paper \\
H$_2$O $M8$ & 23 13 $45.365$  &  61 28 $10.32$ & 0.05 & 0.04 &  This paper \\	
\hline 
H$_2$O $M6$ & 23 13 $45.351$  &  61 28 $10.39$ & 0.05 & --- & This paper \\
H$_2$CO 4.8-GHz $II$\tablenotemark{c} & 23 13 $45.351$ & 61 28 $10.37$ & 0.05 & 0.02 & Hoffman et al. (2003) \\
\enddata
\tablenotetext{a}{Offset from the source listed first}
\tablenotetext{b}{Average position of the $Ia$ and $Ib$ components reported in Table 4 of H03.}
\tablenotetext{c}{Position extracted from Figure 5 of H03.}
\end{deluxetable}

\newpage

\begin{figure}
\begin{center}
\includegraphics[angle=-90,scale=0.7]{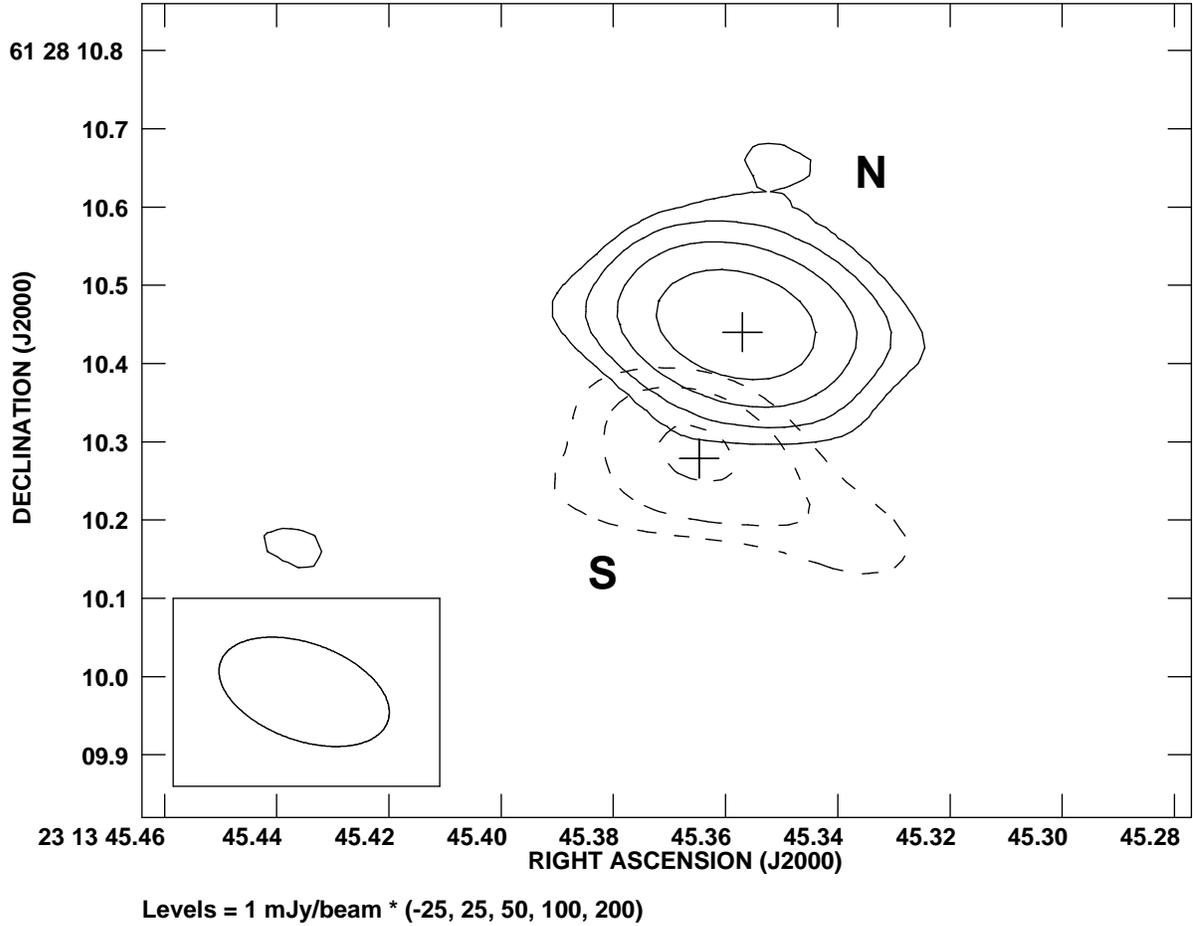} 
  \caption{Contour plot of the 23.1-GHz CH$_3$OH emission toward NGC~7538 IRS~1.
   The solid contours show the North  component while the dashed contours show
   the South component.  The crosses indicate the peak positions from 2-D Gaussian
   fits. Both sets of contours, and their respective crosses, were obtained from the 
   peak channel of the corresponding maser. 
   The synthesized beam is shown in the lower left corner.
   The image rms in each channel is 7~mJy~beam$^{-1}$. }
  \label{fig:f1}
\end{center}
\end{figure}

\newpage

\begin{figure}
\begin{center}
\includegraphics[angle=0,scale=1.0]{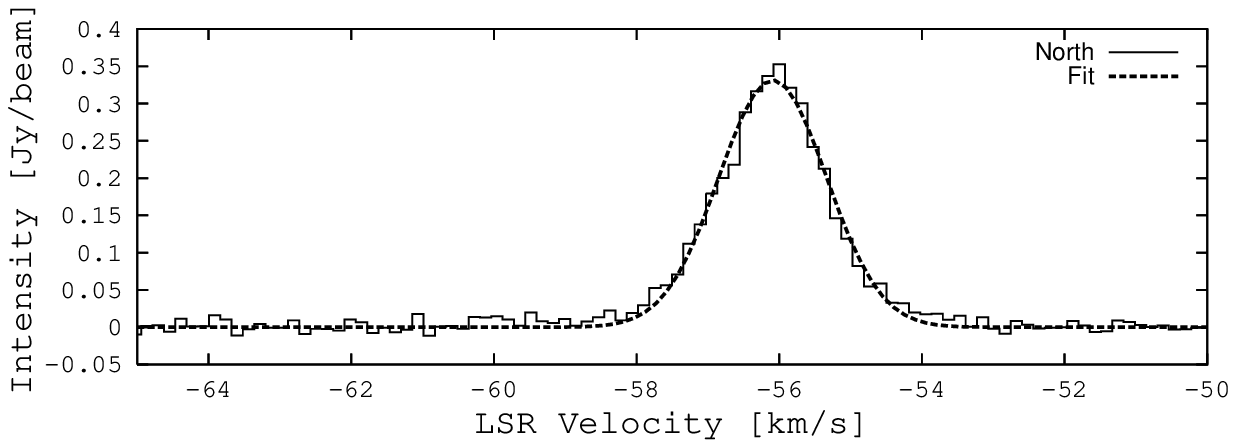}
\includegraphics[angle=0,scale=1.0]{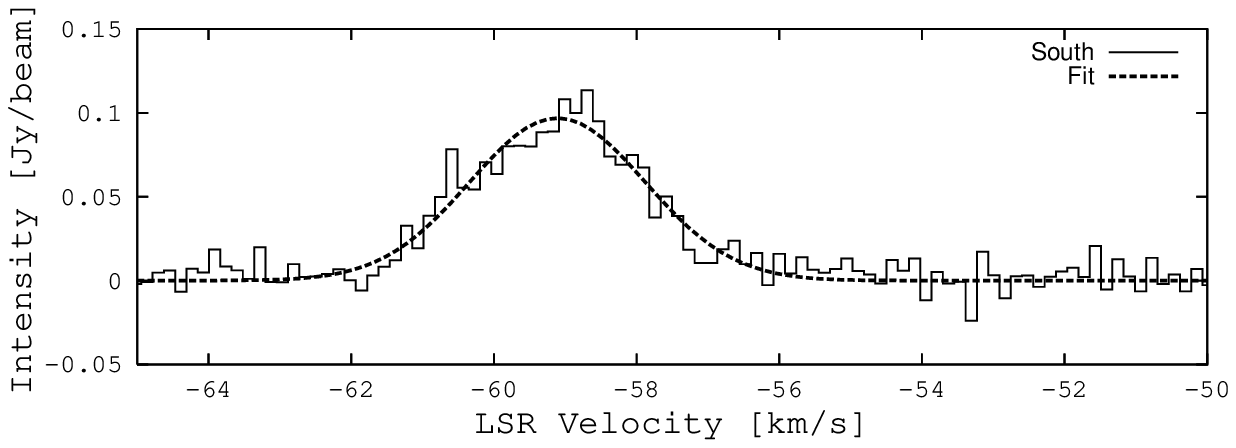}
\caption{CH$_3$OH $9_2-10_1$ $A\tothe +$ spectra at 23.1~GHz toward the NGC 7538 
IRS~1 region. 
\textit{Top:} Spectrum toward the northern maser {\it N} (see \S 3.1). 
\textit{Bottom:} Spectrum toward the southern maser {\it S}.
Gaussian fits are plotted with dashed lines.}
\end{center}
\label{fig:spec}
\end{figure}

\newpage

\begin{figure}
\begin{center}
  \plottwo{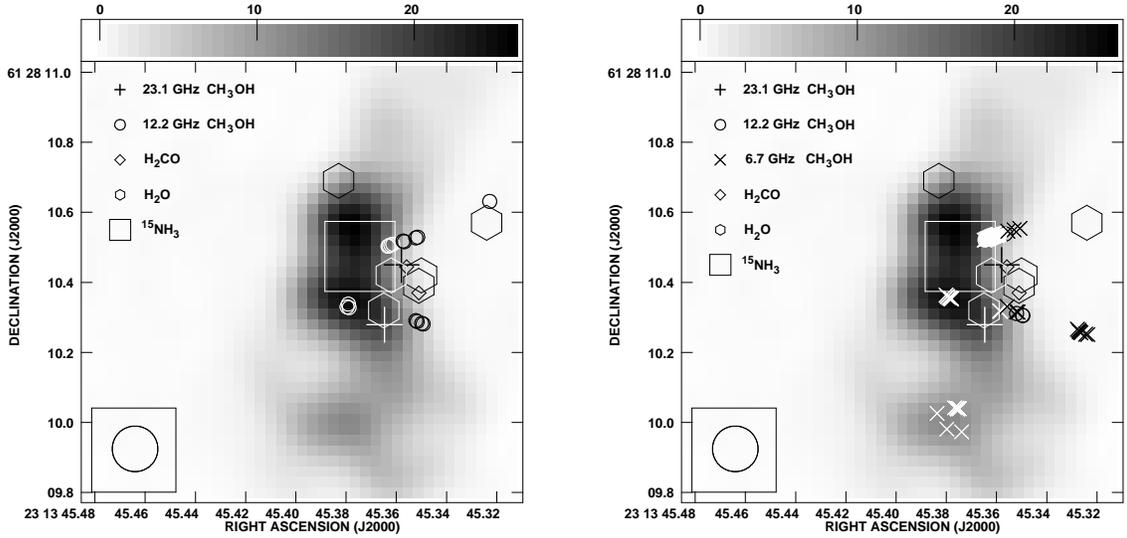}{f3b.eps}
  \caption{
  Overlay of the 2-cm continuum ({\it gray scale}) 
  with masers ({\it symbols})
  toward NGC 7538 IRS~1. 
  The {\it left} panel shows the 12.2-GHz CH$_3$OH masers ({\it circles}) reported by 
  \cite{Mosca09}. The {\it right} panel shows the 6.7-GHz ($\times$ symbols) and 12.2-GHz 
  ({\it circles}) CH$_3$OH masers reported by \cite{Minier00}. Both panels are identical in the 
  rest of their content. 
  The {\it square} at the core of the nebula
  denotes the region with multiple $\tothe{15}$NH$_3$ masers
  reported by \cite{Gaume91}. The {\it hexagons} mark the 22.2-GHz H$_2$O 
  masers reported in this paper.
  The {\it diamonds} mark the 4.8-GHz H$_2$CO
  masers observed by \cite{Hoff03}. 
  The {\it crosses} mark the two 23.1-GHz CH$_3$OH masers reported in this paper. 
  The 2-cm continuum
  image is from \cite{Ramiro04}; its synthesized
  beam is shown in the lower left corner. 
  The symbol size of the new masers reported in this paper  
  corresponds to the upper limit to their absolute position uncertainty $(\pm0.05\arcsec)$. 
  The symbol size for the masers from \cite{Hoff03}, \cite{Minier00}, and \cite{Mosca09} 
  is larger than their absolute position 
  uncertainty (as small as $\pm0.001\arcsec$ for M09). The absolute position uncertainty of 
  the background continuum is similar to that of the new masers we report. The relative uncertainties 
  for a given observation are smaller than the symbols for all cases. 
  }
  \label{fig:f2}
\end{center}
\end{figure}

\end{document}